\documentclass[conference]{IEEEtran}
\IEEEoverridecommandlockouts

\usepackage[left=1.62cm,right=1.62cm,top=1.9cm]{geometry}
\usepackage{cite}
\usepackage{amsmath,amssymb,amsfonts}
\usepackage{algorithmic}
\usepackage{graphicx}
\usepackage{textcomp}
\usepackage{xcolor}
\usepackage{array}
\usepackage{subcaption}
\def\BibTeX{{\rm B\kern-.05em{\sc i\kern-.025em b}\kern-.08em
    T\kern-.1667em\lower.7ex\hbox{E}\kern-.125emX}}
\begin{document}

\title{Effect of Hotspot Traffic on Blocking Probability in Elastic Optical Networks}

\author{\IEEEauthorblockN{Paresh Upadhyay and Yatindra Nath Singh }
\IEEEauthorblockA{\textit{Department of Electrical Engineering, Indian Institute of Technology Kanpur, Kanpur, India} \\
pareshupadhyay2018@gmail.com and ynsingh69@gmail.com }

}

\maketitle

\begin{abstract}
In a circuit-switched network, traffic can be characterized by several factors that define how communication resources are allocated and utilized during a connection. The amount of traffic basically determines how frequently connection requests arrive, how long the setup connection remains active, and the bandwidth used. The Poisson Arrival Process models traffic arrival events at random intervals. It assumes that events happen independently of one another. This model is ideal for simulating traffic in networks where arrivals happen independently and randomly, such as the start of phone calls, data requests, or packet transmissions. The Poisson Arrival Process and uniformly choosing source and destination pair is been used most commonly by researchers to generate traffic in a network to test various promising routing and spectrum assignment algorithms. It checks the algorithm in uniformly loaded conditions and estimate its baseline performance. In real real-world scenario, a bunch of network nodes can start experiencing heavy data traffic compared to the rest of the network. This can lead to latency issues, or even outages if the network is not optimized to handle the load at these nodes which are also called hotspots. In other terms, hotspot in a network is an area or set of nodes within the network that have a higher likelihood of being involved in communication or data transmission compared to other areas. In this paper, we have tried to find what are the various factors involved in increasing the blocking probability in hotspot traffic scenarios. We have also compared the results with the uniform traffic load conditions in same topology.

\end{abstract}


\section{Introduction}

As a significant advancement in optical networking technology, Elastic Optical Networks (EONs) have emerged to overcome the drawbacks of conventional fixed-grid Wavelength Division Multiplexing (WDM) networks. EONs make use of flexible spectrum allocation techniques to dynamically modify the bandwidth of optical channels in accordance with the incoming demands. This allows for more effective use of the optical spectrum. Advanced technology including reconfigurable optical add-drop multiplexers (ROADMs), flexible transponders, and complex network control algorithms are used to achieve this versatility. ITU-T G.694.1 \cite{b1} specifies 12.5 GHz bandwidth slots as the fundamental allocation unit for EONs. For a connection, an integral multiple of 12.5GHz can be requested. \cite{b2} has examined the issue of the optimal basic slot width and concluded that either 6.25 GHz or 12.5 GHz should be used. EONs has two major constraints which need to be satisfied while allocating resources to a request. First one is Spectrum Continuity Constraint which says that a single optical connection (lightpath) must use the same frequency spectrum slots along the entire path from the source to the destination. Second constraint is Spectrum Contiguity which says that the allocated spectrum for a connection must be a contiguous group of spectrum slots \cite{b3}. 

There can be many real life Scenario leading to Hotspots in a network; some of them are as follows. Multiple data centers in a metropolitan area experience simultaneous demand for inter-data center communication (such as during peak business hours or batch processing times); the optical links between these data centers might face localized congestion. In the event of a natural disaster or large-scale emergency, a high number of communication sessions can be initiated from specific geographic regions, such as the affected area or emergency response centers. For instance, hospitals, police stations, and rescue teams may require large amounts of bandwidth for real-time data transfer, video feeds, or coordination. During major events, such as tech conferences, sports events, or concerts, a large number of attendees in a confined location (like a stadium or conference center) may require high-bandwidth services, such as live streaming, video calls, or file-sharing. The optical circuits at the backbone may become congested due to the sudden surge in demand for lightpaths within small zones. If all available wavelengths (optical circuits) are used up, new requests for bandwidth or connectivity may be blocked, leading to increased congestion.

Uniform random traffic is generally used by researchers to test their algorithms in network. Hotspot traffic can affect the working of the algorithm and may cause a decrease in their efficiency. We need to analyze the effect of Hotspot traffic and various other aspects which lead to an increase in blocking probability. The analysis will also be helpful for network operators to better utilize the resources. Earlier \cite{b4},\cite{b5} investigated the highest possible multicast packet throughput on packet-switching WDM ring networks with a hotspot transporting broadcast, multicast, and unicast traffic.     

The paper is organized as follows. We have formulated the routing and spectrum assignment problem for both uniform and hotspot traffic in section II. In section III, We have shown, what are the different ways one can describe the importance of a node in a network. In section IV,  we have created different hotspot scenarios and describe in detail what factor leads to an increase or decrease in blocking probability. In section V, we finally conclude our paper. 

\section{Routing and Spectrum assignment problem (RSA) in EONs}

We will be using NSFNET topology to understand the impact of different traffic scenarios on blocking probability. NSFNET has 14 nodes and 21 bidirectional links as shown in Fig.1. Let NSFNET be represented using a graph $G(V, E)$ having set of nodes, $V$, interconnected using set of links, $E$. Nodes are switches in optical networks and optical fibers are the links that connect the switches. Each edge has 320 spectrum slices, each having a 12.5 GHz width. The length of a link $i$ is $L^i$  where $i  \in E$. The shortest path is found using the Dijkstra method \cite{b6} between $s-d$ pair and resources are allocated using first-fit spectrum assignment method \cite{b7}. Spectrum is assigned while satisfying contiguity and continuity constraints. The required number of spectrum slots $SS_{req}$ is determined using equation (1) \cite{b8}.  
 
\begin{equation}
SS_{req} =  \left\lceil{\frac{b}{W}}\right\rceil
\label{ChoiceM}
\end{equation}

$W$ is the slot width which is 12.5 GHz. To prevent interference, a single-sided guard band of 12.5 GHz has also been taken into consideration between neighboring connections on the same fiber. At each node, request $req(s, d, b)$ will be generated using Poison distribution having an arrival rate of $\lambda$. Here, $s$ is the source node, $d$ is the destination node and $b$ denotes the required bandwidth randomly decided, which follows uniform distribution from 50 Gz to 250 Gz. In uniform traffic, $\lambda$ will be same at each node, and the probability of selecting any two nodes as $s-d$ pair will be same. The holding time of any request follows exponential distribution with mean $1/\mu$. Here $\mu$ is the average service rate of excepted requests. The average traffic load at each node is ($\lambda/\mu$) which will be the same as the average network load $NL$ in case of uniform traffic.

Let us consider $\lambda_{HP}$ and $\lambda_{NHP}$ as the arrival rates at the hotspot nodes and non-hotspot nodes, respectively. For uniform traffic case $\lambda_{HP}$ will be equal to $\lambda_{NHP}$. For Hotspot traffic case $\lambda_{HP}$ will be greater than $\lambda_{NHP}$. Let $\mu$ be the same at each node. To keep the $NL$ same, we need to keep the sum of arrivals rates at all the nodes equal to $\mu*NL$. We have varied $NL$ from 5 to 25 Erlangs. Two different ways to generate hotspot traffic have been considered. One is to keep lambda different for hotspot and non-hotspot nodes as described above and the other way is to keep the higher probability for selecting the hotspot nodes as the destination node while keeping lambda same at each node. The first way will create the scenario where few of the nodes will start setting up connections at a higher rate than others, and the second way will create the condition where few of the nodes will be destination more frequently than other nodes.  We have used two performance metrics namely request blocking probability(RBP) and bandwidth blocking probability (BBP) as described below.    

$$RBP = \frac{\text{Number of blocked requests}}
{\text{Total number of requests}}$$

$$BBP = \frac{\text{Total bandwidth blocked}}
{\text{Total bandwidth requested}}$$

\begin{figure}[ht] 
  \centering
  \includegraphics[width=.92\linewidth,height=.3\textheight]{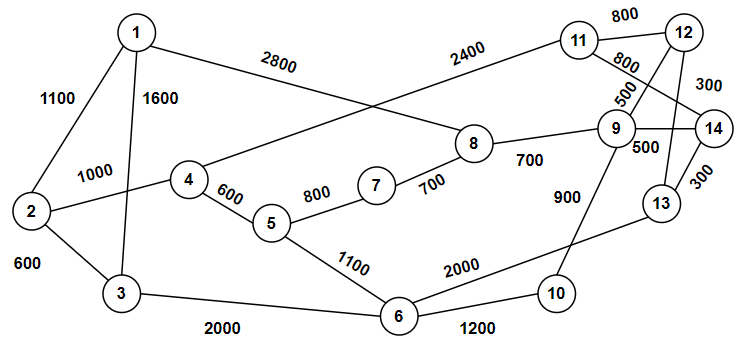}
  \caption{NSFNET}
  \label{NSFNET}
\end{figure}
\section{Different ways to describe importance of a node in the network}

In a network, the importance of a node can be described through various metrics and properties that measure how critical a node is to the network's structure, communication efficiency, and functionality. Different methods emphasize different aspects of the network, such as the node's position, connectivity, or its role in maintaining overall network stability.

Here are the most common ways to describe the importance of a node in a network \cite{b9}.

1. Degree Centrality: Measures the number of direct connections (edges) a node has to other nodes. Nodes with a higher degree are more "connected" and thus expected to have a more significant role in communications within the network. For a node
$i$, its degree centrality $C_d(i)$
is describe as
deg($i$),
where deg($i$) is the number of edges connected to node i. Nodes with high degree are directly involved with more nodes and can be key in disseminating information or serving as hubs for local traffic.

2. Node Betweenness Centrality (NBC): Measures how often a node appears on the shortest paths between other pairs of nodes. Nodes with high betweenness are critical for routing, and control the flow of information in the network. For a node
i, betweenness centrality $C_b(i)$ is describe in equation (2).

\begin{equation}
C_b(i) =  \sum_{s\neq i \neq t} \frac{\sigma_{st}(i)}{\sigma_{st}}
\label{ChoiceM}
\end{equation}

where $\sigma_{st}$ is the total number of shortest paths between nodes s and
t, and $\sigma_{st}(i)$ is the number of those paths that pass through node i. Nodes with high betweenness are vital for connecting different parts of the network, and their removal can severely disrupt communication between other nodes

There are many other metrics like Eigenvector Centrality, PageRank Centrality, Closeness Centrality, etc. We have studied the impact of making nodes with higher and lower node Betweenness Centrality as a hotspot node to blocking probability. Node numbers with there corresponding NBC value for NSFNET are shown in Table 1. 

\begin{table}[ht]
\caption{Nodes with thier corresponding NBC for NSFNET}
\begin{center}
\begin{tabular}{ | m{3em} | m{.5cm}| m{.5cm} | m{.5cm} | m{.5cm} | m{.5cm} | m{.5cm} | m{.5cm} | m{.5cm} | m{.5cm} | m{.5cm} | m{.5cm} | m{.5cm} | m{.5cm} | m{.5cm} | } 
  \hline
  Node no. & 1 & 2 & 3 & 4 & 5 & 6 & 7  \\ 
  \hline
   NBC  &0.14 &0.24 &0.18 &0.31 &0.28 &0.23 &0.26 \\ 
  \hline
  Node no. &8 & 9 & 10 & 11 & 12 & 13 & 14 \\ 
  \hline
  NBC  & 0.34 &0.37 &0.17 &0.21 &0.30 &0.21 &0.14 \\ 
  \hline
  
\end{tabular}
\end{center}
\end{table}

\section{Network performance under different Traffic condition} 

To generate the hotspot traffic we have changed the distribution of selecting the destination node from uniform to non-uniform. 
We have selected 30\% of nodes that is four nodes in  NSFNET, which will act as a hotspot. These four nodes will be selected based on node betweenness centrality. We have selected four nodes with the lowest centrality values (nodes 1,3,10 and 14) and in another simulation four nodes with the highest centrality values (nodes 4,8,9 and 12) as hotspot nodes. The probability of selecting a hotspot node as destination node is .2 and .02 for selecting a non-hotspot node. The requests will be generated at the rate of $\lambda$ per second using Poisson distribution. $\lambda$ is kept same for all the nodes.

 When RSA is run on uniform traffic, its performance in a network is indicated using SP-FF. Performance for shortest path routing with first-fit spectrum assignment when run on NSFNET making nodes 1,3,10 and 14 as hotspot nodes, is indicated using SP-FF-HP1. Shortest path routing with first-fit spectrum assignment when run on NSFNET making nodes 4,8,9 and 12 as hotspot nodes, its performs is indicated using SP-FF-HP2. The simulation was run five times for one lakh CRs, and the average of these runs was used to determine the outcome. Continuity and Contiguity constraints are satisfied while assigning the spectrum slots.

RBP performance of the SS-FF, SS-FF-HP1, and SP-FF-HP2 is shown in Fig.2.a. SS-FF-HP1 gives 24.5\% more RBP when averaged over all the loads compared to the SS-FF algorithm which uses uniform traffic condition for same average traffic. SS-FF-HP2 gives 62.4\% more RBP when averaged over all the loads compared to SS-FF.  SS-FF-HP2 gives 29.6\% more RBP than SS-FF-HP1. It can be seen that the location of the hotspot node can significantly affect the increase or decrease of RBP of a network even after keeping the average load of the network the same. 
Fig.2.b. shows the BBP vs different traffic loads for all three traffic conditions. SS-FF-HP1 gives 25.2\% more BBP when averaged over all load than the SS-FF algorithm. SS-FF-HP2 gives 58.2\% more BBP when compared to then SS-FF.  SS-FF-HP2 gives 25.5\% more BBP when compared to SS-FF-HP1. BBP results indicate different Hotspot node locations can significantly increase the blocking of a number of spectrum slots.

\begin{figure}[ht]
	\centering
	\begin{subfigure}{0.8\linewidth}
	\includegraphics[width=\linewidth,height=.3\textheight]{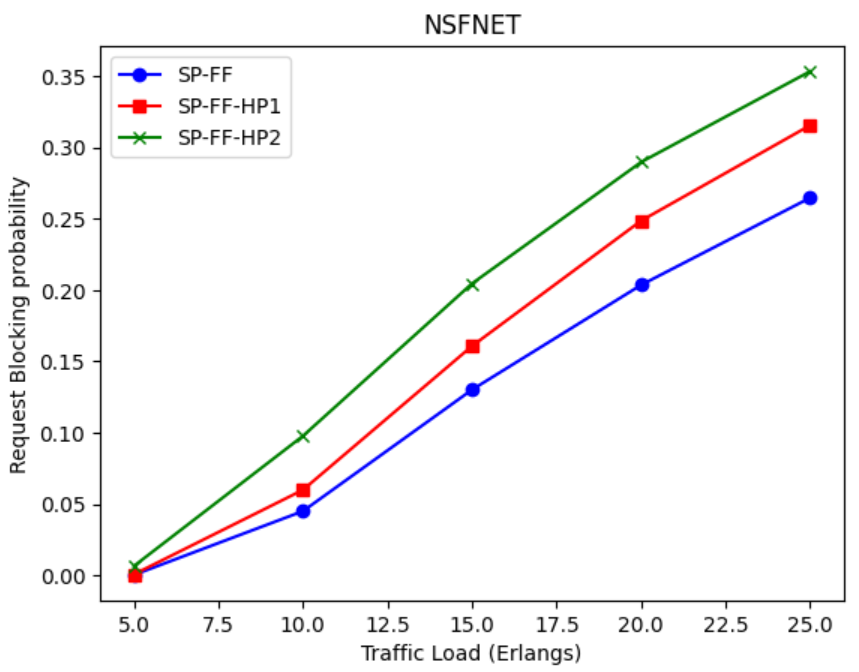}
		\caption{}
		\label{fig:subfigA}
	\end{subfigure}
	\begin{subfigure}{0.8\linewidth}
	\includegraphics[width=\linewidth,height=.3\textheight]{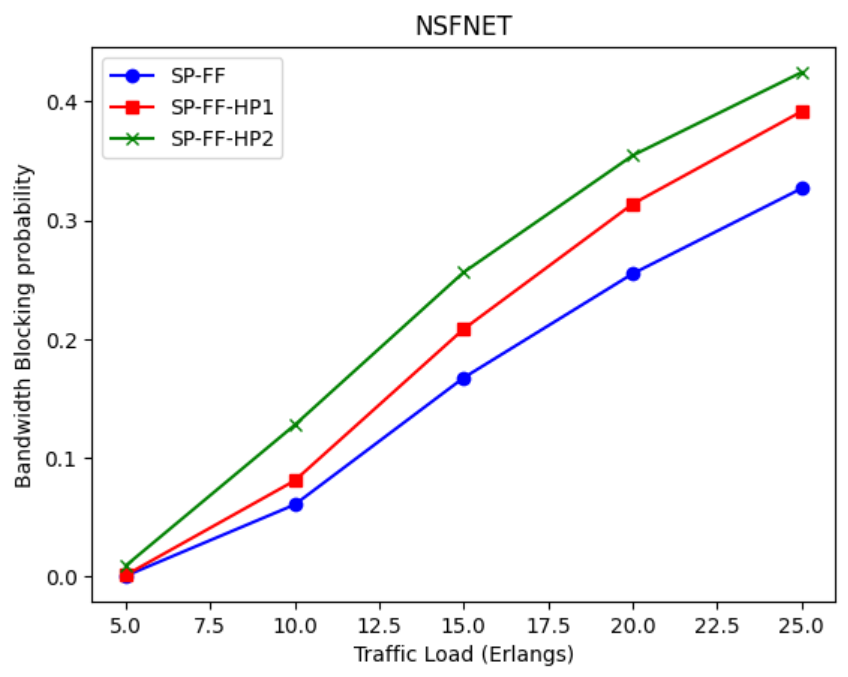}
		\caption{}
		\label{fig:subfigB}
	\end{subfigure}
	\caption{a) RBP b) BBP vs Traffic load in NSFNET }
	\label{fig:subfigures}
\end{figure}
Earlier we have seen the impact of the location of hotspot nodes in increasing blocking probability. Now we will study how the number of hotspot nodes can impact the performance metric of a network. We have kept the total percentage of requests received by the hotspot nodes the same which is 80\%. We will be using two hotspot traffic scenarios having different numbers of hotspot nodes. Firstly we have considered the lowest four nodes according to NBC that is 1,3, 10, and 14, and their performance on different loads is indicated by SP-FF-N4, The probability of selecting a hotspot node is .2 and for selecting a non-hotspot node in .02. Secondly, we have considered six lowest NBC valued node that is 1,3, 10, 11, 13 and 14 as hotspot node and its performance is indicated by SP-FF-N6, probability of choosing hotspot node is .166 and for non-hotspot node is .025.

RBP performance under a different number of hotspot node conditions is shown in Fig.3.a. SP-FF-N6 gives 13.7\% less RBP than SP-FF-N4 which indicates that as the number of hotspots increases by keeping the number of requests received the same, traffic gets more uniformly distributed which decreases the blocking probability.

BBP performance under a different number of hotspot node conditions is shown in Fig.3.b. SP-FF-N6 gives 14\% less BBP than SP-FF-N4 which indicates a decrease in blocking of the number of slots as the number of hotspots increases which again indicates traffic gets more uniformly distributed and a load of specific links decreases. 
\begin{figure}[ht]
	\centering
	\begin{subfigure}{0.8\linewidth}
	\includegraphics[width=\linewidth,height=.3\textheight]{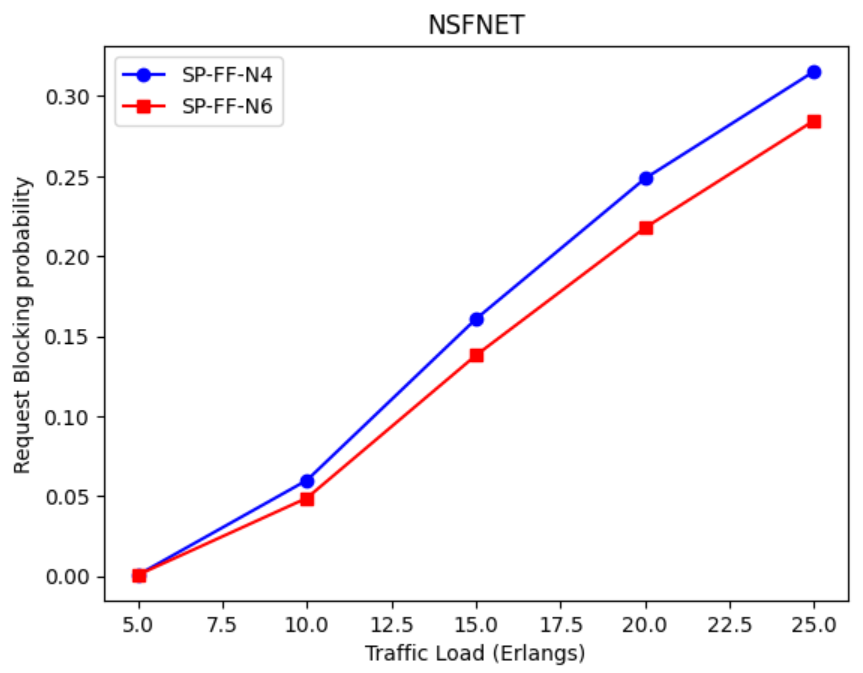}
		\caption{}
		\label{fig:subfigA}
	\end{subfigure}
	\begin{subfigure}{0.8\linewidth}
	\includegraphics[width=\linewidth,height=.3\textheight]{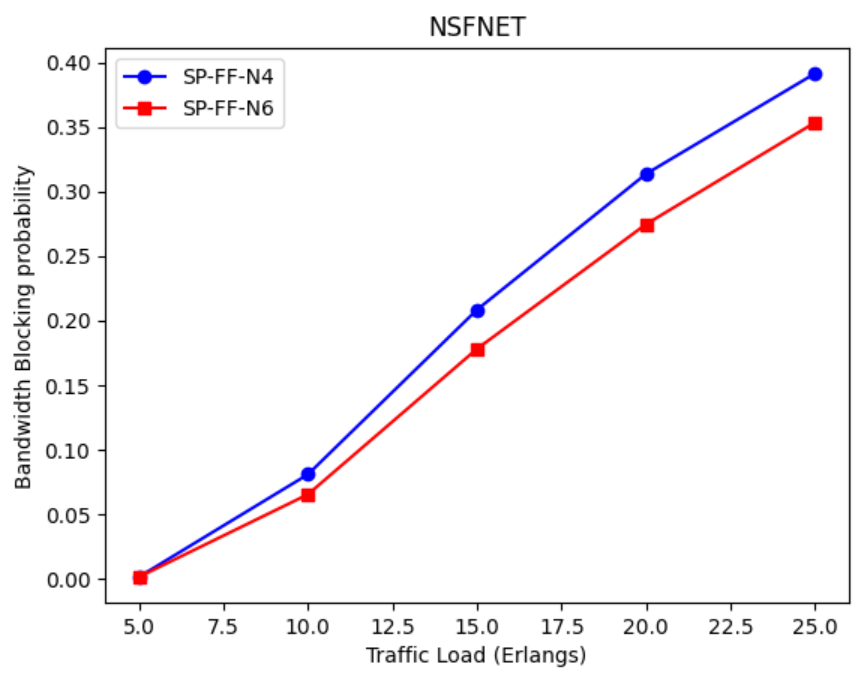}
		\caption{}
		\label{fig:subfigB}
	\end{subfigure}
	\caption{a) RBP b) BBP vs Traffic load in NSFNET }
	\label{fig:subfigures}
\end{figure}

Let us consider the scenario where the generation rate of requests is kept high at certain nodes (Hotspot nodes) by keeping the average traffic load of a network the same. We have considered nodes 1, 3, 10, and 14 hotspot nodes. Let us consider $\lambda_{NHN}$  as the rate of generation of requests at the non-hotspot node and $\alpha*\lambda_{HN}$ at the hotspot node. We need to keep the sum of all the rates the same for all $\alpha$ to keep the average load of the network the same which lies in the range from 5 to 25 Erlang. The probability of selecting any $s-d$ pair is kept equal. Fig.4.a and Fig.4.b shows the RBP and BBP of various loads for $\alpha$ equal 1, 2, 4 and 6. SS-FF indicates the shortest path algorithm applied when the rate of generation of request at each node is $\lambda$. SS-FF-2lambda indicates the application of the shortest path algorithm when the rate of generation of request at hotspot nodes is $2*\lambda$ and so on, it should be noted that the rate of generation of request is decreased at non-hotspot nodes to keep the average network load the same.  

RBP value when averaged over all load shows no significant increase for SS-FF-2lambda but for SS-FF-4lambda there is an increase of 5\% and for SS-FF-6lambda there is an increase of 15\% when compared to SS-FF. This indicates when a few nodes start to transmit more data as compared to other nodes there is an increase in blocking of requests. BBP when averaged over all load also shows no significant increase for SS-FF-2lambda but for SS-FF-4lambda there is an increase of 6\% and for SS-FF-6lambda there is an increase of 16\% when compared to SS-FF. This also indicate an increase in blocking due to the overloading of few links. 

\begin{figure}[ht]
	\centering
	\begin{subfigure}{0.8\linewidth}
	\includegraphics[width=\linewidth,height=.3\textheight]{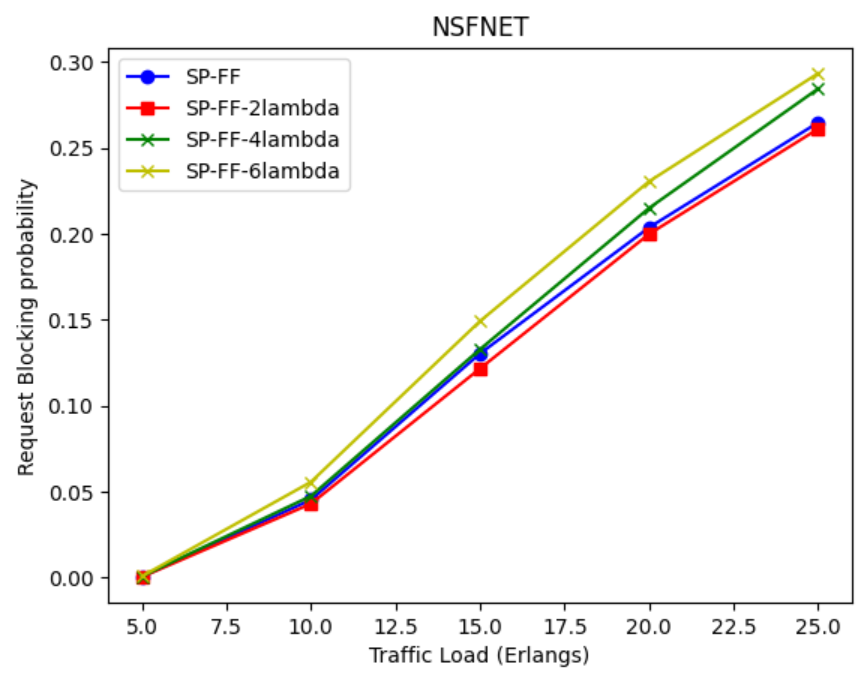}
		\caption{}
		\label{fig:subfigA}
	\end{subfigure}
	\begin{subfigure}{0.8\linewidth}
	\includegraphics[width=\linewidth,height=.3\textheight]{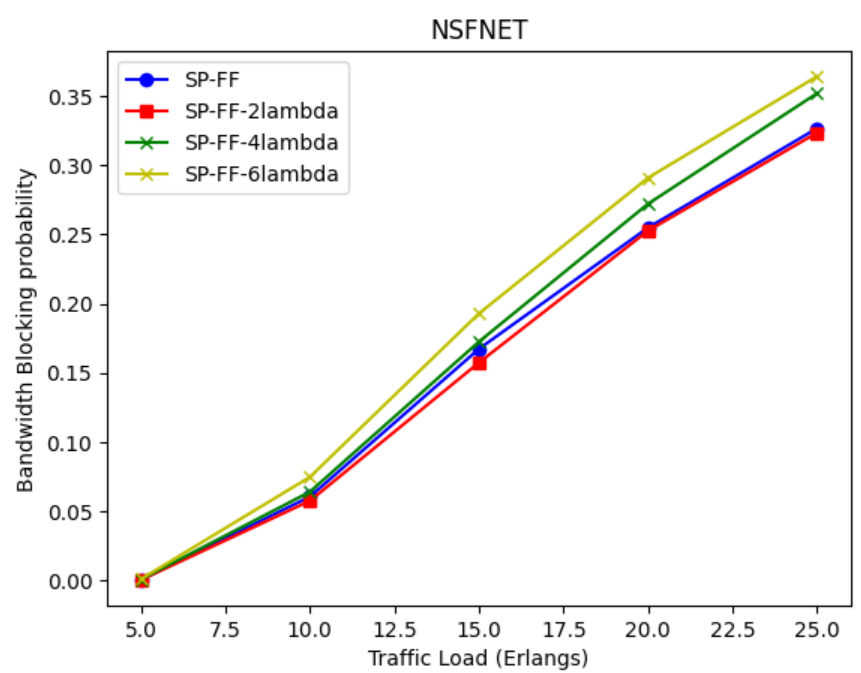}
		\caption{}
		\label{fig:subfigB}
	\end{subfigure}
	\caption{a) RBP b) BBP vs Traffic load in NSFNET }
	\label{fig:subfigures}
\end{figure}

\begin{figure}[ht]
	\centering
	\begin{subfigure}{0.8\linewidth}
	\includegraphics[width=\linewidth,height=.3\textheight]{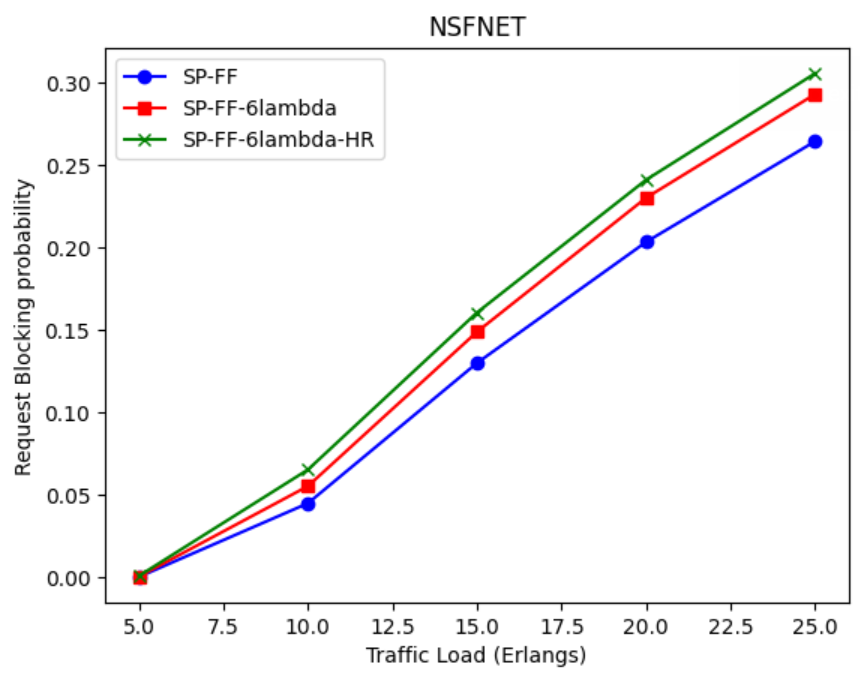}
		\caption{}
		\label{fig:subfigA}
	\end{subfigure}
	\begin{subfigure}{0.8\linewidth}
	\includegraphics[width=\linewidth,height=.3\textheight]{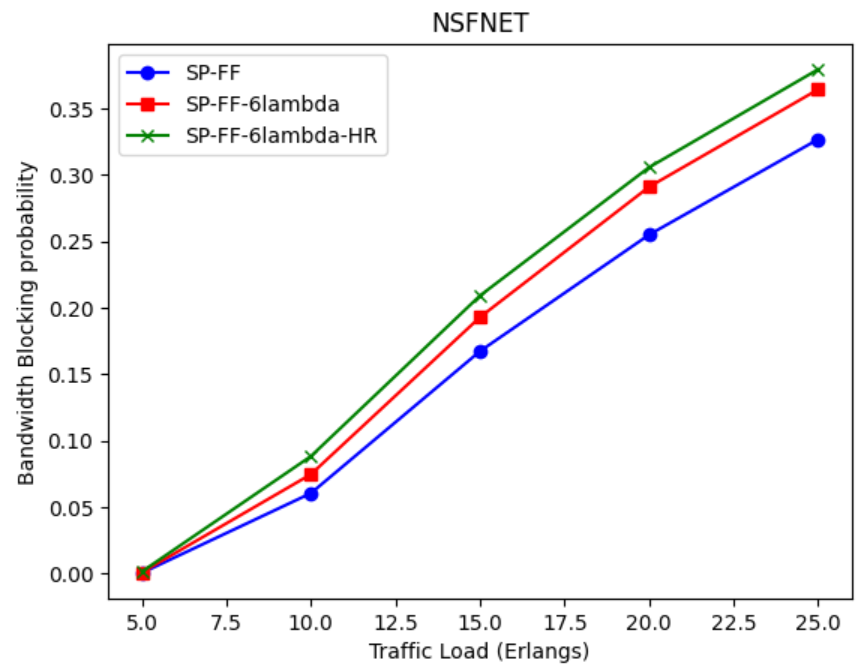}
		\caption{}
		\label{fig:subfigB}
	\end{subfigure}
	\caption{a) RBP b) BBP vs Traffic load in NSFNET }
	\label{fig:subfigures}
\end{figure}

In our last case study, we have created a closed hotspot region which means few of the nodes start to transmit data among themselves with high data rates. We will consider nodes 1, 3, 10, and 14 for creating closed hotspot regions. Whenever a source node belongs to a hotspot node it will select a node from a set of hotspot nodes with probability 0.2 and rest with probability .02. We will also consider the rate of generation of request at hotspot nodes to be 6 times that of the not-hotspot node. RBP and BBP performance of closed hotspot region indicated using SS-FF-6lambda-HR.

Fig.5.a. and Fig.5.b. shows the performance of all three scenarios. There is an increase in 25.7\% and 8.7\% of RBP of SS-FF-6lambda-HR from SS-FF and SS-FF-6lambda. Formation of closed hotspot region also increases BBP by 26.7\% and 9\% when compared to SS-FF and SS-FF-6lambda. These results simply indicate that the formation of a closed hotspot region will overload a few links which will make it difficult for new connections to satisfy both the constraints.

\section{CONCLUSION}
In this paper, we have studied what are the different factors responsible for increasing our decreasing blocking probability in hotspot traffic scenarios and also compared the results with blocking probability calculated in uniform traffic scenarios. We have studied the impact of hotspot nodes location according to node NBC value and found out that if nodes with higher NBC value start behaving as hotspots will significantly increase the blocking probability. We have also studied the impact of several hotspot nodes on the RBP metrics by keeping the total percentage of requests received the same by changing the probability of selecting the hotspot and non-hotspot node. We have found that as the number of hotspot nodes increases RBP goes down indicating a more uniform distribution of load across the network. We have also considered the scenario where the rate of generation of requests is kept high at Hotspot nodes by keeping the average traffic load of a network the same, this also results in increased blocking probability. A closed hotspot region is also formed and results show further increase in RBP and BBP when compared to all the cases. Present of hotspot nodes in general will increase the blocking probability when compared to a uniform traffic scenario.

\end{document}